\documentclass{egpubl}
\usepackage{pg2025s}

\SpecialIssuePaper         

\CGFStandardLicense

\usepackage[T1]{fontenc}
\usepackage{dfadobe}  

\usepackage{cite}  
\BibtexOrBiblatex
\electronicVersion
\PrintedOrElectronic
\ifpdf \usepackage[pdftex]{graphicx} \pdfcompresslevel=9
\else \usepackage[dvips]{graphicx} \fi

\usepackage{egweblnk} 
\usepackage{lineno}
\usepackage{amsmath}
\usepackage{amssymb}

\usepackage{xspace}

\newcommand{\eg}{{\emph{e.g.}},\xspace}

\title{SPG: Style-Prompting Guidance for Style-Specific Content Creation }

\author[Q. Liang et al.]{
	\parbox{\textwidth}{\centering Qian Liang \qquad Zichong Chen \qquad Yang Zhou\thanks{Corresponding author.}  \qquad Hui Huang \\
		\parbox{\textwidth}{\centering Visual Computing Research Center, CSSE, Shenzhen University }
	}
}

\begin{document}
	
	\teaser{
		\includegraphics[width=\linewidth]{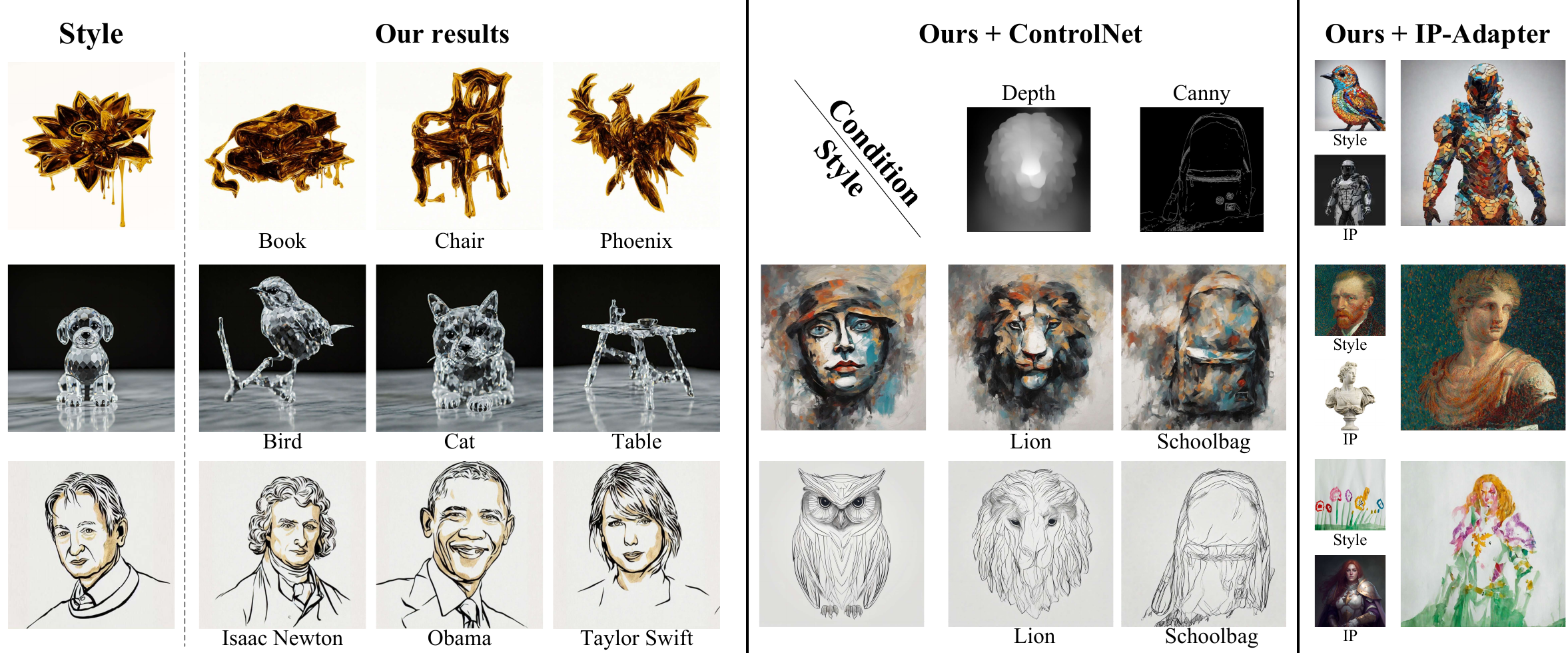}
		\centering
		\caption{ We propose a training-free method that is capable of generating stylized images according to a given style reference (left). The generated images not only align with the visual characteristics of the style example but also adhere to the semantics of the input text prompt. As a sampling method, our approach can be easily integrated with many existing diffusion plugins, such as ControlNet, to enable more fine-grained control over the generated stylized images using conditions like depth maps or Canny edges (middle), or IP-Adapter by replacing the text prompts with image conditions as the content guidance (right).} 
		\label{fig:teaser}
	}
	\maketitle
	
	\begin{abstract}
		Although recent text-to-image (T2I) diffusion models excel at aligning generated images with textual prompts, controlling the visual style of the output remains a challenging task. In this work, we propose Style-Prompting Guidance (SPG), a novel sampling strategy for style-specific image generation. SPG constructs a style noise vector and leverages its directional deviation from unconditional noise to guide the diffusion process toward the target style distribution.  By integrating SPG with Classifier-Free Guidance (CFG), our method achieves both semantic fidelity and style consistency. SPG is simple, robust, and compatible with controllable frameworks like ControlNet and IPAdapter, making it practical and widely applicable. Extensive experiments demonstrate the effectiveness and generality of our approach compared to state-of-the-art methods. Code is available at \url{https://github.com/Rumbling281441/SPG}.

		\ccsdesc[300]{Computing methodologies~Image processing}

		\printccsdesc   
	\end{abstract}  
	

	\section{Introduction}
	
	The field of text-to-image (T2I) generation has witnessed remarkable progress in recent years. With the continuous advancement of pretrained diffusion-based models\cite{ramesh2021zero,rombach2022highlatentdiffusion,saharia2022photorealistic,chang2023muse}, it is now possible to synthesize high-quality images conditioned on textual prompts. However, in scenarios where a reference style image is provided, generating images that faithfully reflect both the target style and the prompt semantics remains a challenging task, even though the emergence of numerous methods \cite{hertz2024stylealign,jeong2024vsp,zhou2025attentiondistillation,chen2025styleblend} for style-specific T2I generation. A key difficulty lies in the inherent trade-off between style consistency and textual alignment: many existing approaches either achieve strong stylization at the cost of accurately representing the text prompt, or maintain semantic fidelity but fail to effectively capture the desired style.
	
	Although many approaches have been proposed to address this problem, achieving an effective balance remains elusive. Existing approaches generally can be grouped into three categories: finetuning-based \cite{textualinversion, ruiz2023dreambooth, alaluf2023neural,frenkel2024b-lora}, pretraining-based \cite{ye2023ipadapter,xing2024csgo,wangstyleadapter,gao2024styleshot,ahn2024dreamstyler}, and training-free \cite{hertz2024stylealign,zhou2025attentiondistillation,jeong2024vsp} approaches, while each has notable limitations. Fine-tuning methods typically yield high-quality stylization but require costly adaptation per style. Pretrained approaches often leverage global style tokens or feature alignment strategies, which may have difficulty capturing subtle or local style properties. Training-free methods are more efficient and flexible, yet frequently suffer from reduced style fidelity or semantic misalignment due to limited style representation capacity. Even recent work such as StyleBlend \cite{chen2025styleblend}, which explicitly tackles the trade-off between style fidelity and text alignment, still relies on fine-tuning for each individual style. Despite effectively enhancing both dimensions, such approaches are impractical for large-scale deployment or real-time applications due to their high computational cost and limited flexibility. These limitations motivate us to develop a more practical and generalizable solution.
	
	In this work, we propose Style-Prompting Guidance (SPG), a training-free, sampling-based framework for style-specific T2I generation. Inspired by the widely adopted Classifier-Free Guidance (CFG) \cite{ho2022classifierfreeguidance} and recent feature injection techniques \cite{hertz2022prompttoprompt,tumanyan2023plugandplay,cao2023masactrl}, SPG extends its formulation to incorporate style-distribution guidance. During the diffusion process, SPG modifies the self-attention layers by replacing the Keys and Values in the unconditional branch with those extracted from a style reference image, which produces a style-conditioned noise vector. By subtracting the original unconditional noise, we obtain a directional style guidance signal that guides the sampling trajectory toward the target style distribution.
	
\begin{figure}[tbp]
		\centering
		\includegraphics[width=\linewidth]{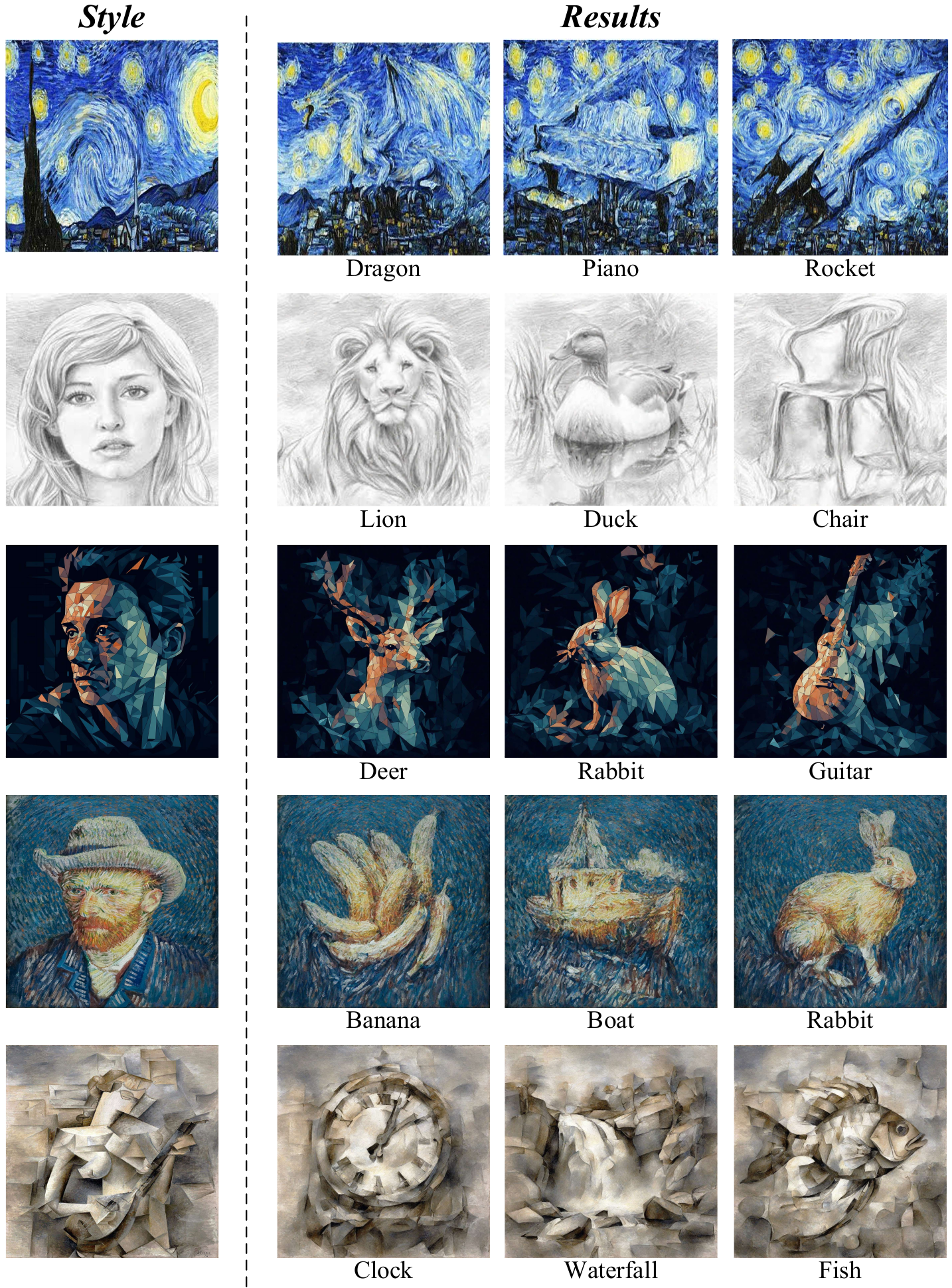}
		\caption{\label{fig1}
			Given a style image and various prompts, our method generates images that simultaneously adhere to the reference style and accurately reflect the prompt content.}
	\end{figure}

	We then treat the stylized generation task as a joint sampling process, in which semantic and style guidance serve as complementary controls. SPG corresponds to enforcing adherence to the reference style, while CFG guarantees the textual semantics. Their combination enables the model to generate images that are semantically accurate and stylistically consistent; see Fig.~\ref{fig1} as an example. Beyond its simplicity and effectiveness, SPG is highly versatile and compatible. As a plug-and-play technique, it can be seamlessly integrated with popular controllable generation frameworks such as ControlNet \cite{zhang2023controlnet} and IP-Adapter \cite{ye2023ipadapter}, and only requires a single reference image for style specification, making it particularly suitable for real applications with limited data or on-the-fly customization requirements; see \eg Fig.~\ref{fig:teaser}.
	
	Experiments across diverse styles and prompts demonstrate the robustness and effectiveness of SPG. Our method consistently outperforms prior approaches in achieving a favorable balance between style fidelity and textual alignment. These results highlight SPG as a practical, robust, and generalizable solution for style-specific T2I generation, offering a compelling combination of stylization strength, semantic faithfulness, and deployment efficiency.

	\section{Related work}
	
	\subsection{T2I diffusion models}
	Text-to-image models\cite{chang2023muse,nichol2021glide,rombach2022highlatentdiffusion} generate high-quality images from given textual descriptions, ushering in a new technological revolution in the field of computer vision. Leveraging their powerful prior knowledge, these models demonstrate strong applicability in tasks such as image editing\cite{couairon2022diffedit,brooks2023instructpix2pix,gu2023photoswap}, inpainting\cite{manukyan2023hd,avrahami2023blended,ohanyan2024zero}, style transfer\cite{chung2024styleid,wang2024instantstyle}, and more.
	
	\subsection{Feature manipulation in diffusion models}
	The diffusion models~\cite{rombach2022highlatentdiffusion,podell2023sdxl} employ a UNet architecture to predict noise for denoising images. These works\cite{hertz2022prompttoprompt,zhou2024self-rectification,sun2025attentive,gu2024swapanything} perform image editing, texture inpainting, and other tasks by manipulating the Query (Q), Key (K), and Value (V) components within the attention blocks of Stable Diffusion.\cite{kwon2022h-space} achieve the disentanglement of semantic information in images by editing the intermediate feature representations within the UNet.
	\cite{cao2023masactrl,tumanyan2023plugandplay,zhou2025attentiondistillation,jeong2024vsp} have empirically demonstrated that the self-attention layers in the UNet encode information about an image’s texture and appearance. Such information can be utilized for style transfer and style image generation. In our work, we also leverage these representations to characterize the style of an image.
	
	\subsection{{Guidance in diffusion sampling}}
	Recent advances in the quality of images generated by Stable Diffusion\cite{rombach2022highlatentdiffusion} can be largely attributed to advancements in sampling guidance techniques\cite{dhariwal2021cg,ho2022classifierfreeguidance,hong2023sag,ahn2024pag}. Classifier guidance\cite{dhariwal2021cg} steers the model toward the distribution defined by the given condition by leveraging the gradient of a classifier's loss, thereby facilitating the generation of high-quality results. CFG \cite{ho2022classifierfreeguidance} eliminates the need for an additional classifier by incorporating unconditional noise during the sampling process, thereby achieving an effect similar to that of classifier guidance.
	\subsection{Style-specific image generation}
	Generating an image that aligns with the style of a given reference image while adhering to a textual prompt is a highly important task with significant practical applications. Existing approaches include both training-based methods and training-free methods. Training-based methods~\cite{textualinversion,ruiz2023dreambooth, Multiconcept,ouyang2025klora,sohn2023styledrop} usually fine-tune the text-to-image model using a given reference style image, enabling the model to internalize the style and subsequently generate images that match both the style and various prompts. However, these approaches are highly resource- and time-intensive, and it is often challenging to obtain multiple images with the same style for fine-tuning a diffusion model. Training-free methods~\cite{hertz2024stylealign,zhou2025attentiondistillation,jeong2024vsp} typically extract style information from a reference image—specifically the K and V from the self-attention layers—and inject them into the generation process of the target prompt, enabling the synthesis of images in the desired style. However, the generated images often fail to accurately preserve either the style information or the semantic content of the prompt. We believe that the injection of K and V in these methods primarily disrupts the CFG sampling process (see the discussion in Section~\ref{constrains} and the comparative analysis in Section~\ref{sec:comparison}). Generating images that simultaneously adhere to both the desired style and the prompt semantics remains highly challenging. 
	\section{Preliminaries}
	\begin{figure}[tbp]
		\centering
		\includegraphics[width=\linewidth]{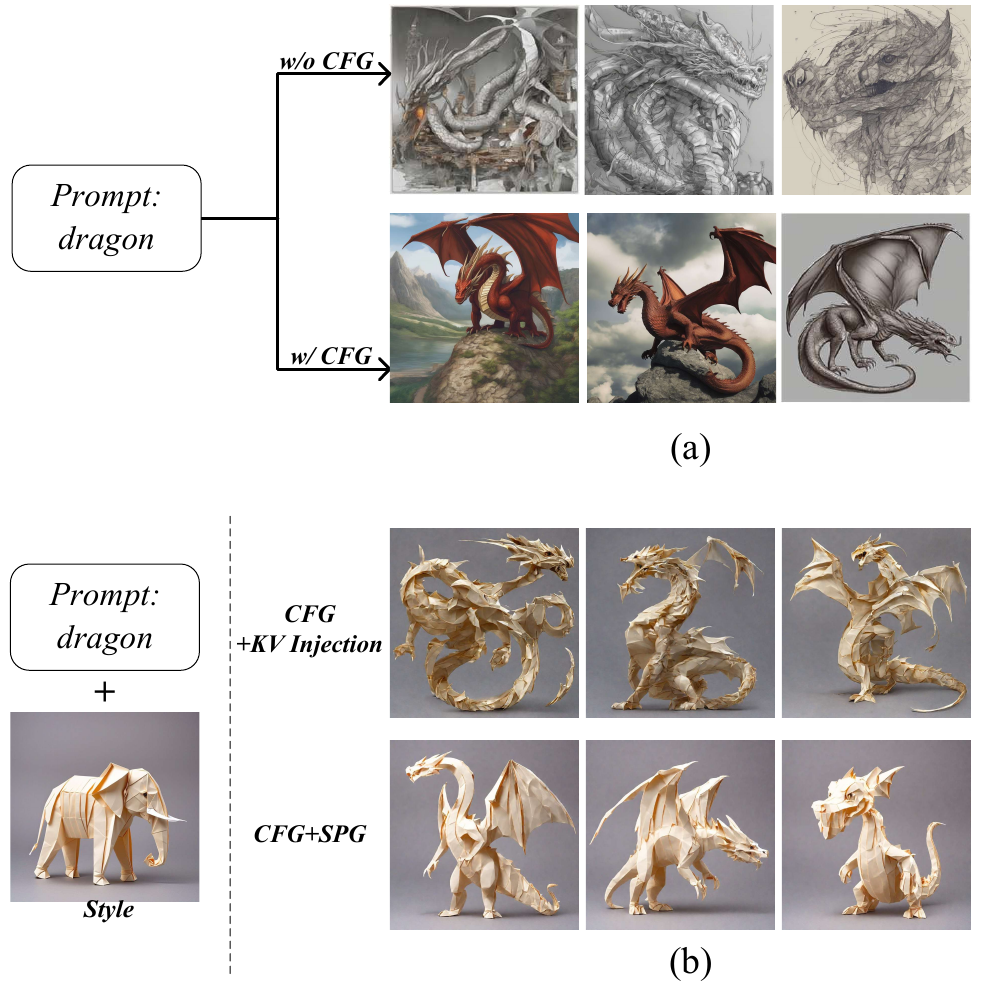}
		\caption{\label{cfg1}
			(a) Given a prompt, e.g., dragon, the first row shows results without using CFG, while the second row shows results with CFG applied. We can see that CFG is crucial in guiding the diffusion process toward the distribution specified by the text prompt. 
			(b) When there's a style reference, directly injecting KV features from the style image into the CFG process of the target image generation cannot guarantee reasonable results (the first row). Instead, this simple injection disrupts the CFG guidance, leading to degraded image quality and poor text alignment. As a comparison, our results (the second row) achieve a plausible balance between style coherence and semantic alignment.}
	\end{figure}

	\subsection{Latent diffusion models}
	Latent diffusion models\cite{rombach2022highlatentdiffusion} are trained in the latent space of images. During training, an input image is first encoded into a latent representation $z$ using a pretrained encoder $\mathcal{E}$. Gaussian noise is progressively added to $z$, and a UNet-based denoising network is trained to predict the noise at each timestep. The training objective minimizes the discrepancy between the true noise $\epsilon\sim\mathcal{N}(0,\textbf{1})$ and the predicted noise $\epsilon_\theta$, conditioned on a textual prompt $c$. The objective is formulated as:
	\begin{equation}
	\mathcal{L}_{\text{LDM}} = \mathbb{E}_{z,\epsilon,t} \left[ \|\epsilon - \epsilon_{\theta}(z_t, t, c)\|_2^2 \right],
	\end{equation}
	where $z_t$ denotes the noisy latent at diffusion timestep $t$, and $\epsilon_\theta$ is the noise prediction generated by the UNet denoiser parameterized by $\theta$, conditioned on the textual embedding $c$. 
	
	In this work, we adopt Stable Diffusion \cite{rombach2022highlatentdiffusion} as our backbone to validate the proposed Style-Prompting Guidance (SPG), taking advantage of its well-explored UNet-based architecture to enable effective style-conditioned noise manipulation during the sampling process.

	\subsection{Feature injection in diffusion models}
	\begin{figure*}[t]
		\centering
		\includegraphics[width=\linewidth]{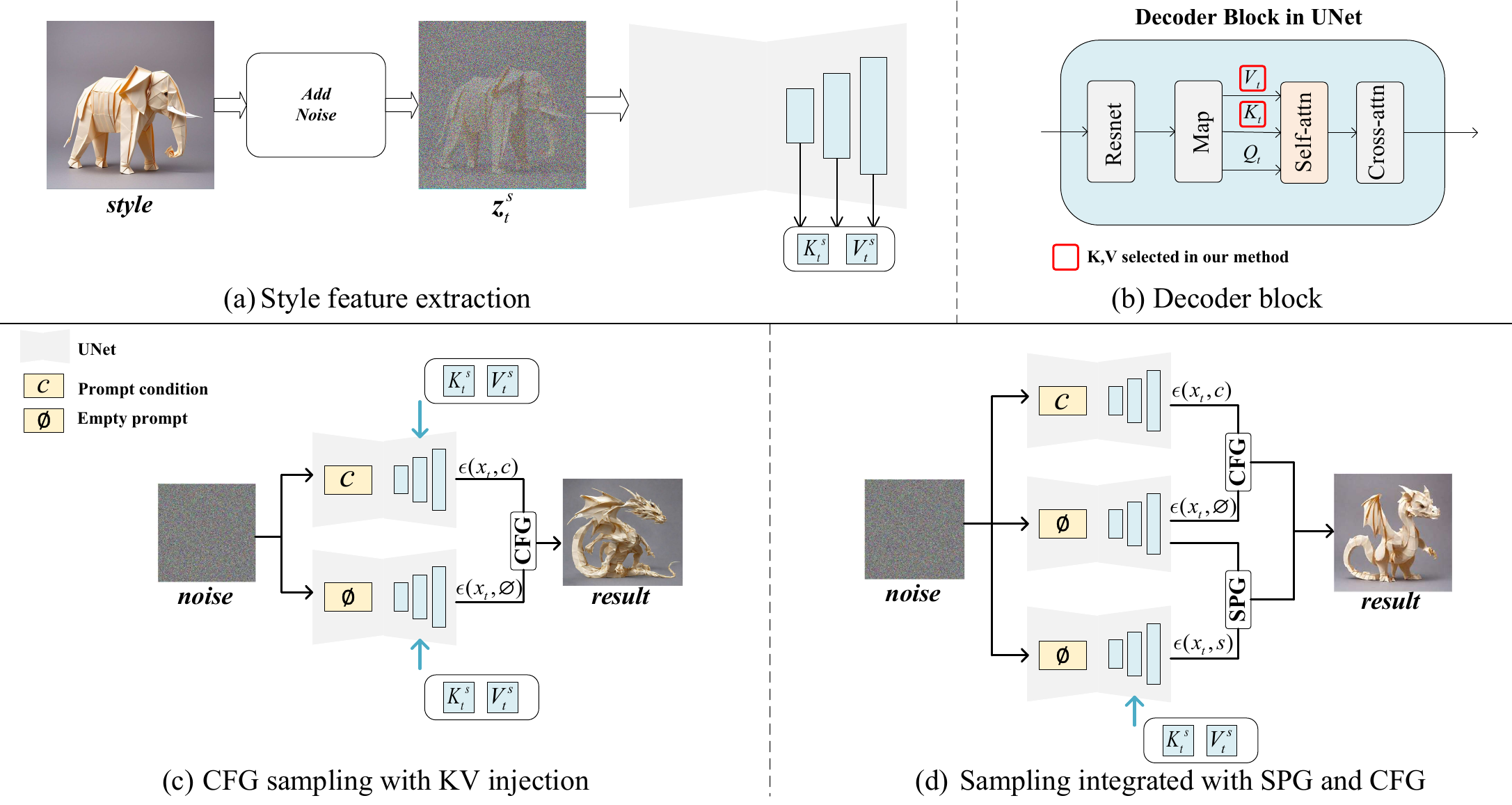}
		\caption{\label{method}\textbf{Method Overview.} (a) To acquire style features of a given reference, we perform DDIM\cite{song2020DDIM} inversion and extract the attention features. Specifically, at each diffusion step $t$, we add noise to the style image (in the latent space) to obtain $z_t^s$, which is then denoised by the U-Net, and the $K_t^s$ and $V_t^s$ features from the self-attention layers of the U-Net decoder, as shown in (b), are regarded as the style representation. (c) A straightforward way to generate new content is to perform CFG sampling with KV injection. At each timestep, the KV features extracted previously are injected into both the conditional (with prompt $c$) and unconditional (with empty prompt $\emptyset$) branches of the CFG, which, however, is proven suboptimal in generating high-quality content. (d) In our method, we have three branches in sampling, one conditional and two unconditional. We only inject the style features into one of the unconditional branches to construct a style-aware conditional noise, while keeping the original CFG unchanged. Subsequently, we perform mixed sampling using both CFG and SPG.}
	\end{figure*}
	
	Recent studies on feature injection in diffusion models have demonstrated that a substantial amount of semantic and stylistic information is encoded within the self-attention layers of the UNet denoising network \cite{hertz2022prompttoprompt,tumanyan2023plugandplay,cao2023masactrl}. In particular, the attention mechanism plays a pivotal role in enabling fine-grained control over both the content and appearance of generated images. Our work specifically focuses on the self-attention modules within the UNet architecture, where the attention operation is formally defined as:
	\begin{equation}
	\text{Self{-}Attention}(Q, K, V) = \text{softmax}\left( \frac{QK^T}{\sqrt{d}} \right) V,
	\end{equation}
	where $Q$, $K$, and $V$ represent the {query}, {key}, and {value} matrices, respectively; $d$ denotes the dimensionality of the key vectors.
	
	Empirical findings~\cite{tumanyan2023plugandplay,cao2023masactrl,hertz2022prompttoprompt} suggest that the Query matrix primarily encodes semantic information, such as spatial structure and object layout, while the Key and Value matrices are more closely associated with visual appearance attributes, including texture, color, and style. Building on this insight, feature injection methods~\cite{hertz2024stylealign,zhou2025attentiondistillation,jeong2024vsp} typically extract the $KV$ components from a style reference image and inject them into the self-attention computation of the target image during the generation process. This technique facilitates effective style transfer without requiring any model parameter updates, offering a flexible and efficient approach for controllable image synthesis.

	\subsection{Classifier-free guidance}
	Classifier-Free Guidance (CFG) \cite{ho2022classifierfreeguidance} is a widely used sampling strategy in diffusion models to improve alignment with textual prompts. An example of CFG can be seen in Fig.~\ref{cfg1}(a). The guided noise prediction is computed as:
	\begin{equation}
	\tilde\epsilon(x_t,c) = \epsilon(x_t, \emptyset) + \lambda_{\text{CFG}} \cdot \left( \epsilon(x_t,c) - \epsilon(x_t,\emptyset) \right),
	\label{cfg}
	\end{equation}
	where $\tilde\epsilon(x_t,c)$ denotes the guided noise prediction, $\epsilon(x_t,c)$ is the noise estimated by the model conditioned on the prompt $c$, and $\epsilon(x_t,\emptyset)$ represents noise conditioned on an empty text prompt.
	
	CFG can be viewed as shifting the sampling trajectory toward the semantic distribution defined by the text prompt, effectively guiding the generation toward content that better matches user intent.
	
	\section{Method Overview}
	Given a single style reference image and a text as the prompt, our goal is to mimic the target style and synthesize images using text-to-image diffusion models, ensuring consistency with the reference style while preserving the semantics of the prompt.
	Previous training-free approaches~\cite{hertz2024stylealign,zhou2025attentiondistillation,jeong2024vsp} attempt to achieve this by injecting the KV features of the style images into the self-attention layers during CFG sampling. However, we observe that this often leads to style degradation or semantic alignment, ultimately resulting in suboptimal stylization quality. To better understand these limitations, we first construct a KV injection baseline that directly incorporates style features into the CFG process, and analyze its shortcomings in Sec.~\ref{constrains}. Motivated by these observations, we then present our proposed Style-Prompting Guidance (SPG), a novel mechanism that preserves the integrity of the CFG path while enabling strong style control. We detail the construction of SPG in Sec.~\ref{4.2}, and its integration with CFG within a unified sampling framework in Sec.~\ref{sec:spgcfg}.

	\subsection{KV injection baseline}\label{constrains}
	
	To synthesize images in a specific style, KV injection approaches primarily involve two steps: i) extracting KV features from the self-attention layers of a style image, and ii) injecting these KV features into the generation process of the target image.
	
	\textbf{Style feature extraction.} To ensure computational efficiency, we adopt the feature extraction procedure proposed in~\cite{jeong2024vsp}, as illustrated in Fig.~\ref{method}(a).
	At each diffusion timestep $t$, we first apply forward diffusion to the style image to match the noise level of the current step. We then use a pretrained denoising UNet to extract KV features (denoted as $K_t^{s}$ and $V_t^{s}$) from its self-attention layers, as depicted in Fig.~\ref{method}(b). These features encode texture and appearance information and serve as the style representation for guiding the subsequent image generation.

	\textbf{KV injection with CFG sampling.} To inject the extracted style KV features into the generation process, we follow a dual-branch CFG strategy: one conditioned on the text prompt and one unconditional. As shown in Fig.~\ref{method}(c), we replace the original self-attention KV features in both branches with $K_t^{s}$ and $V_t^{s}$, thereby injecting the reference style into the denoising process.
	\begin{figure}[tbp]
		\centering
		\includegraphics[width=\linewidth]{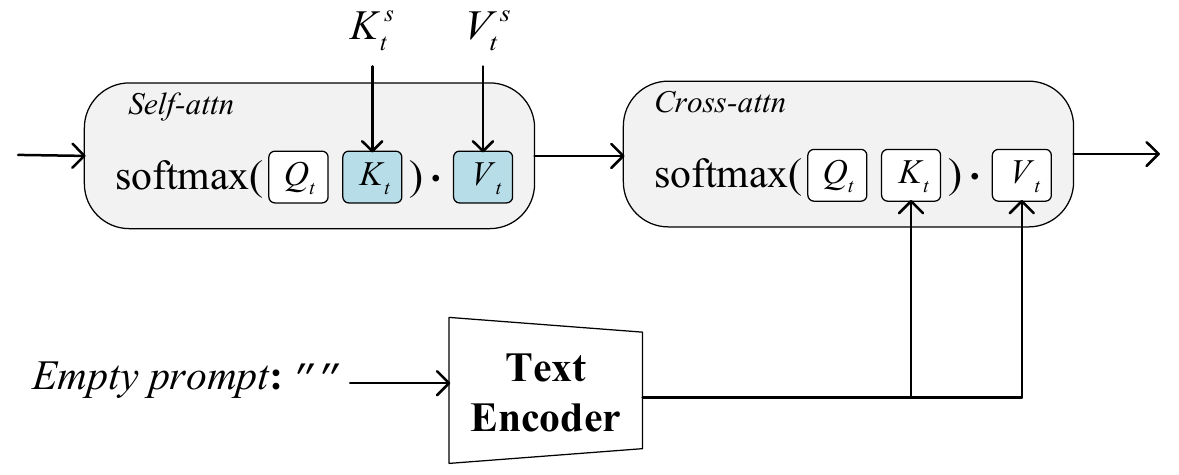}
		\caption{To construct the style-conditioned noise, we replace the attention features ($K_t$ and $V_t$) in the unconditional forward diffusion path with those extracted from the style reference ($K_t^s$ and $V_t^s$).} 
		\label{spg construction}
	\end{figure}
	
	\textbf{Discussion.} While this baseline effectively injects style information, it introduces critical limitations. As demonstrated in Fig.~\ref{cfg1}(a), CFG is essential for ensuring semantic fidelity by leveraging the contrast between the conditional and unconditional noise predictions. However, we observe that directly injecting the style KV features into both branches of the CFG process often leads to weak and inconsistent stylization effects, as shown in Fig.~\ref{cfg1}(b). This degradation arises because the injected KV features are entangled with the original attention context, and the model lacks an explicit mechanism to separate and enhance style-specific attributes. As a result, the generated outputs may fail to faithfully reproduce the desired visual characteristics of the reference image. Also, since the same style features are injected into both the conditional and unconditional branches, the model cannot distinguish style guidance from semantic steering, leading to unstable or inconsistent results.
	
	\begin{figure}[t]
		\centering
		\includegraphics[width=\linewidth]{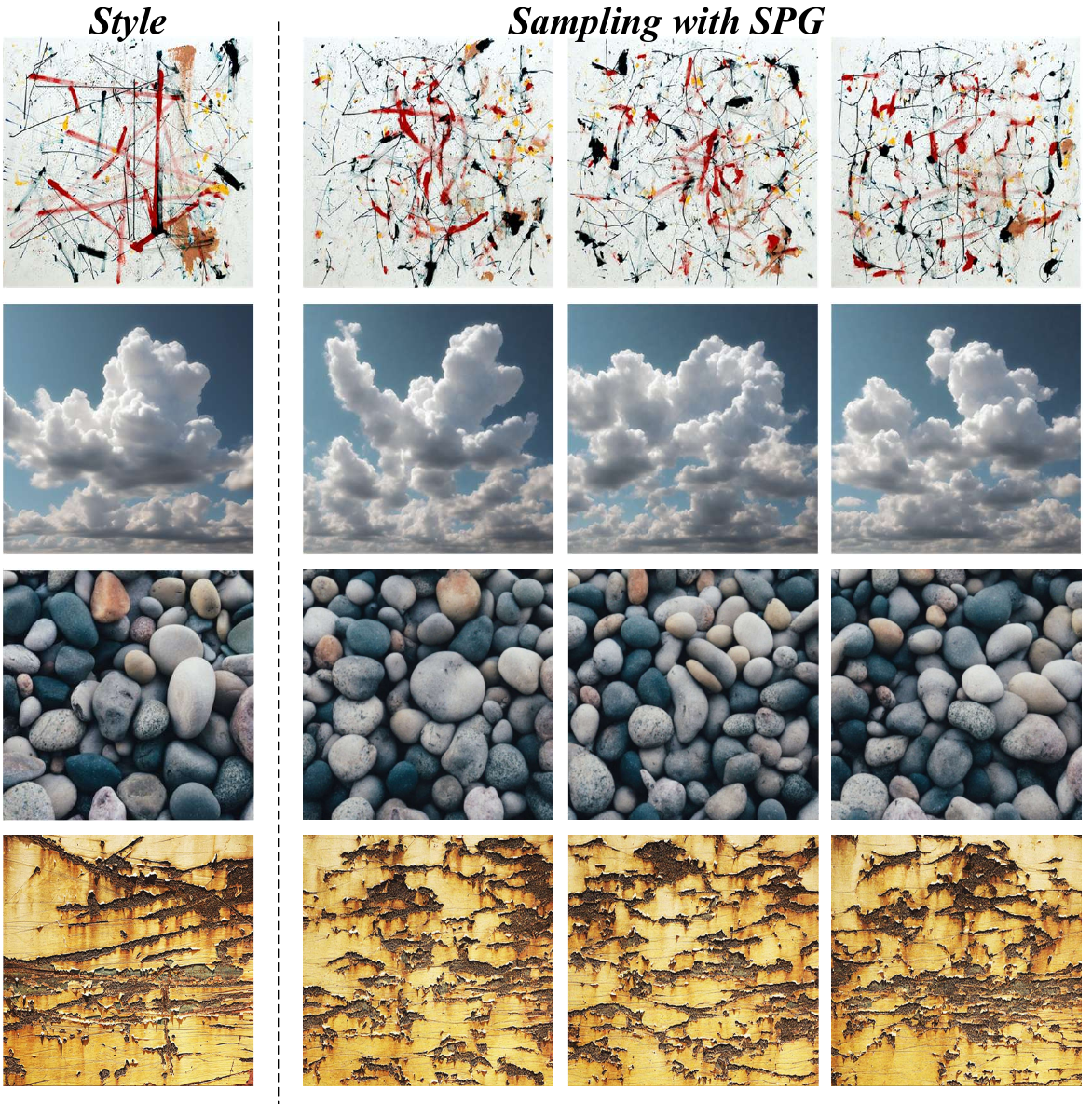}
		\caption{Sampling with SPG can produce image variations that
			faithfully adhere to the appearance of the reference image.}
		\label{spg sample}
	\end{figure}
	
	\begin{figure*}[t]
		\centering
		\includegraphics[width=\linewidth]{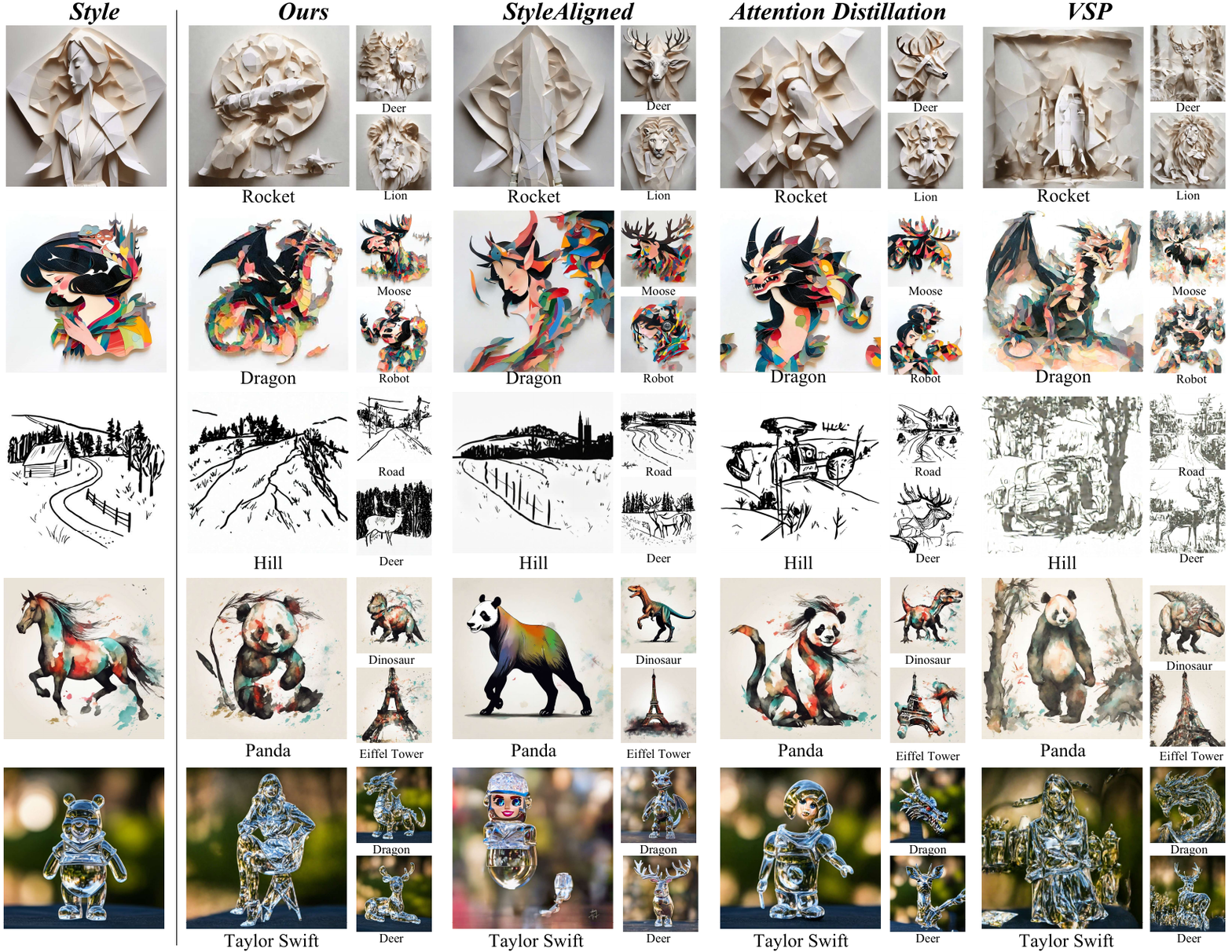}
		\caption{A qualitative comparison with other training-free methods based on KV injection.}
		\label{comparison}
	\end{figure*}
	
	\subsection{Style-prompting guidance}\label{4.2}
	Rather than injecting style features directly into both branches of the CFG framework, we aim to construct a style-conditioned noise signal in a parallel path that complements CFG without interfering with its core mechanism. This forms the basis of our proposed Style-Prompting Guidance (SPG). SPG follows a similar principle of CFG to enable style control. At each diffusion timestep $t$, we first extract the style features $K_t^{s}$ and $V_t^{s}$ from the reference image as described in Section~\ref{constrains}. We then inject them into a new unconditional forward path of the denoising UNet, as illustrated in Fig.~\ref{spg construction}. This forms a style-aware noise prediction $\epsilon(x_t, s)$ that incorporates texture and appearance information from the reference image, while remaining disentangled from text prompt semantics. Formally, SPG is defined as:
	\begin{equation}
	\tilde\epsilon(x_t, s) = \epsilon(x_t, \emptyset) + \lambda_{\text{SPG}} \cdot \left( \epsilon(x_t, s) - \epsilon(x_t, \emptyset) \right),
	\label{spg}
	\end{equation}
	where $\tilde\epsilon(x_t, s)$ denotes the style-guided noise prediction, $\epsilon(x_t, \emptyset)$ is the unconditional noise prediction, and $\lambda_{\text{SPG}}$ controls the strength of the style influence.
	
	To further improve low-level style consistency, particularly in terms of color tone, contrast, and spatial statistics, we additionally apply AdaIN \cite{huang2017adain} at each timestep to align the noisy latents of the target and reference images. Specifically, we apply AdaIN as:
	\begin{equation}
	\text{AdaIN}({z_t^{tgt}}, {z_t^{s}}) = \sigma({z_t^{s}}) \left( \frac{{z_t^{tgt}} - \mu({z_t^{tgt}})}{\sigma({z_t^{tgt}})} \right) + \mu({z_t^{s}}),
	\label{eq:adain}
	\end{equation}
	where ${z_t^{tgt}}$ and ${z_t^{s}}$ are the intermediate latents of the generated image and reference style image, respectively, and $\mu(\cdot)$ and $\sigma(\cdot)$ denote the mean and standard deviation.
	
	SPG guides the model toward the distribution of the style image during the diffusion process by leveraging the semantic difference between a style-conditioned noise and an unconditional noise. Several sampling results obtained using SPG are shown in Fig.~\ref{spg sample}.

	\subsection{Integration of SPG and CFG}\label{sec:spgcfg}
	To generate images that align with both the textual prompt and the reference style, we integrate SPG and CFG into a unified sampling framework, as illustrated in Fig.~\ref{method}(d), where the final noise prediction at each timestep is defined as:
	\begin{equation}
	\begin{split}
	\tilde\epsilon_\theta(x_t,c,s) &= \epsilon_\theta(x_t,\emptyset) + \lambda_{\text{CFG}} \cdot \left( \epsilon_\theta(x_t,c) - \epsilon_\theta(x_t,\emptyset) \right)\\ 
	& + \lambda_{\text{SPG}} \cdot \left( \epsilon_\theta(x_t,s) - \epsilon_\theta(x_t,\emptyset) \right),
	\end{split}
	\end{equation}
	where the term $\lambda_{\text{CFG}}$ and $\lambda_{\text{SPG}}$ control the strength of prompt alignment and style alignment, respectively. This hybrid formulation allows the diffusion process to be guided independently but simultaneously toward the semantic and style distributions. Importantly, since SPG is disentangled from the CFG path, it preserves the contrastive structure critical for semantic alignment. Some results generated by our method are shown in Figs.~\ref{fig:teaser} and ~\ref{fig1}.
	
	\begin{figure}[tbp]
		\centering
		\includegraphics[width=\linewidth]{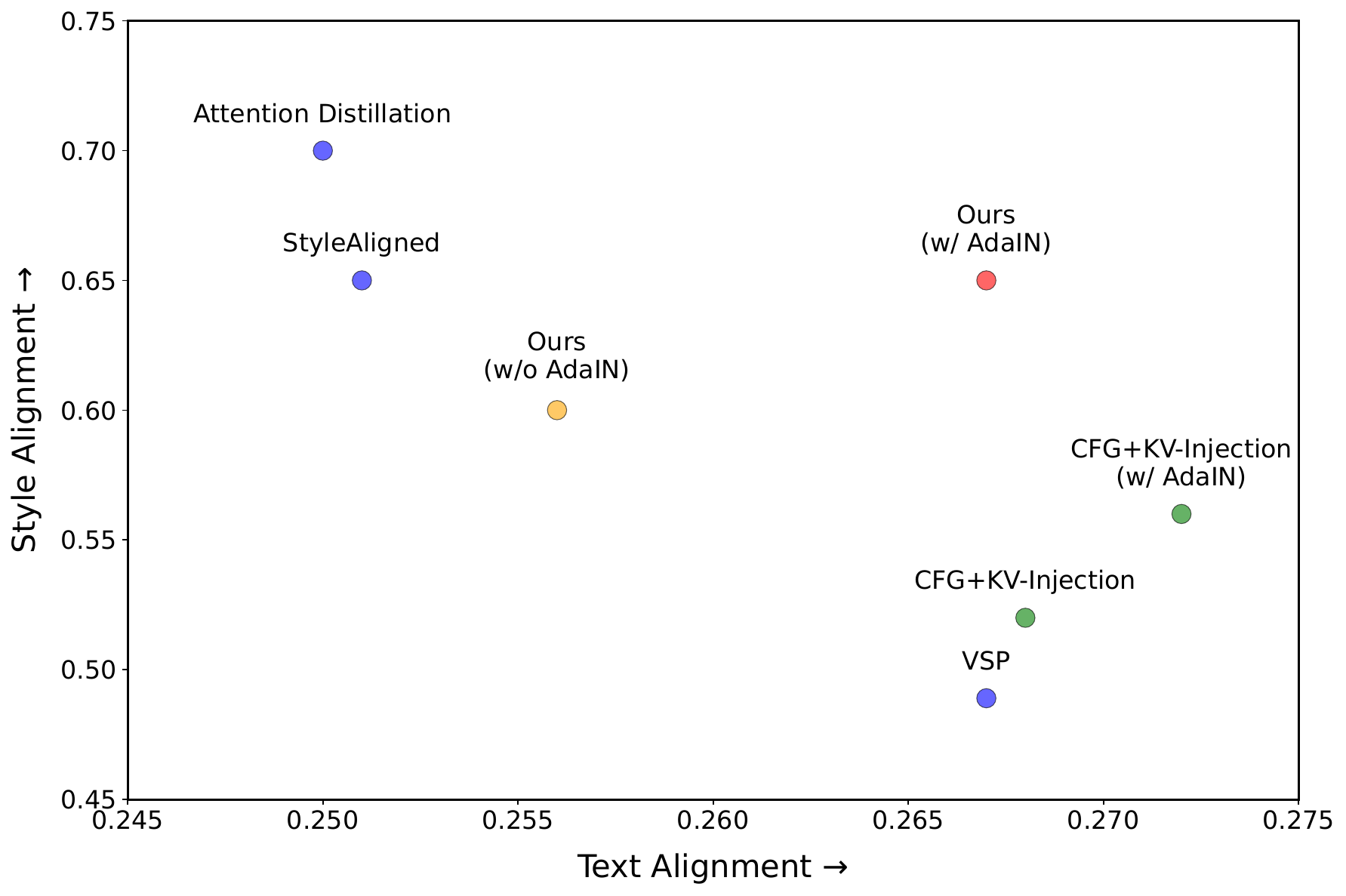}
		\caption{\label{point}Quantitative comparison. We compare our method (red) with various existing methods (blue), our ablated version (orange), and the constructed baselines (green) in terms of text alignment and style alignment. Our method achieves the best balance. }
	\end{figure}

	\section{Experiments}
	\subsection{Implementation details}
	We implement our method on Stable Diffusion XL (SDXL)\cite{podell2023sdxl}. We perform KV feature extraction and injection on the last 6 self-attention layers (layers 65 to 70) of the UNet, following recent experiences~\cite{cao2023masactrl, tumanyan2023plugandplay, jeong2024vsp}. We set the $\lambda_{\text{CFG}}$ to 7.0 and the $\lambda_{\text{SPG}}$ to 3.0 for all experiments, using DDIM\cite{song2020DDIM} sampling with 50 steps. Our method takes approximately 30 seconds to generate an image with a specific style on a single RTX 3090 GPU.

	\textbf{Dataset.} We have collected a set of style images from the internet. For each style image, we generate a set of images using text prompts consisting of words such as \{guitar, deer, chair, balls\}. 
	
	\textbf{Metrics.} 
	To verify whether the generated images correspond to the given prompts, we assess text alignment using CLIP\cite{radford2021clip} cosine similarity. Specifically, we employ the CLIP ViT-B/32 model to extract features from both the generated images and the provided prompts, and then compute the cosine similarity between each image’s features and those of its corresponding prompt to measure text alignment.
	To evaluate style alignment, we extract DINO~\cite{caron2021dino} VIT-B/8 embeddings for both the generated stylized images and the reference style images, and compute the cosine similarity between these embeddings.

	\subsection{Comparisons}\label{sec:comparison}
	To demonstrate the effectiveness of our method, we conduct both qualitative and quantitative comparisons with several KV injection-based approaches. Section~\ref{4.2} as part of the SPG process. 
	Fig.~\ref{comparison} shows the qualitative comparison, where we compare our method with StyleAligned~\cite{hertz2024stylealign}, Attention Distillation~\cite{zhou2025attentiondistillation}, and VSP~\cite{jeong2024vsp}, using the official open-source implementations provided by the authors. As we can see,
	StyleAligned~\cite{hertz2024stylealign} exhibits significant deficiencies in semanticn alignment. This is primarily because StyleAligned injects both the KV information and the Q features of the style image during the CFG-based diffusion process. As analyzed previously (Section~\ref{constrains}), such injection disrupts the semantic distinction between the conditional and unconditional noise, undermining the mechanism of CFG and leading to degraded generation performance. Attention distillation~\cite{zhou2025attentiondistillation} optimizes the KV loss between the target and style latents at each diffusion step $t$ in image generation. Although it does not directly inject KV information into the target generation process during CFG, the latent representations are inherently dependent on both the conditional and unconditional noise. Therefore, this is equivalent to implicitly injecting the KV information of the style image into the target image generation process. As a result, this method also exhibits suboptimal semantic alignment in our comparison.
	VSP~\cite{jeong2024vsp} concatenates KV features of the style image into the generated target during the CFG process, which also leads to both weak image quality and style alignment.

	Fig.~\ref{point} further shows the quantitative comparison with the above competitors, where we use 10 styles and 5 prompts, and generate 5 images per prompt to compute the average statistics.
	In addition, we include the baseline constructed by directly injecting KV feature during the CFG process. To enable a more comprehensive comparison, we further improve the baseline by applying AdaIN at each diffusion time step between the generated target latent and the corresponding reference latent. Our method achieves the best balance between style fidelity and text alignment among all test methods.

	\begin{figure}[t]
		\centering
		\includegraphics[width=\linewidth]{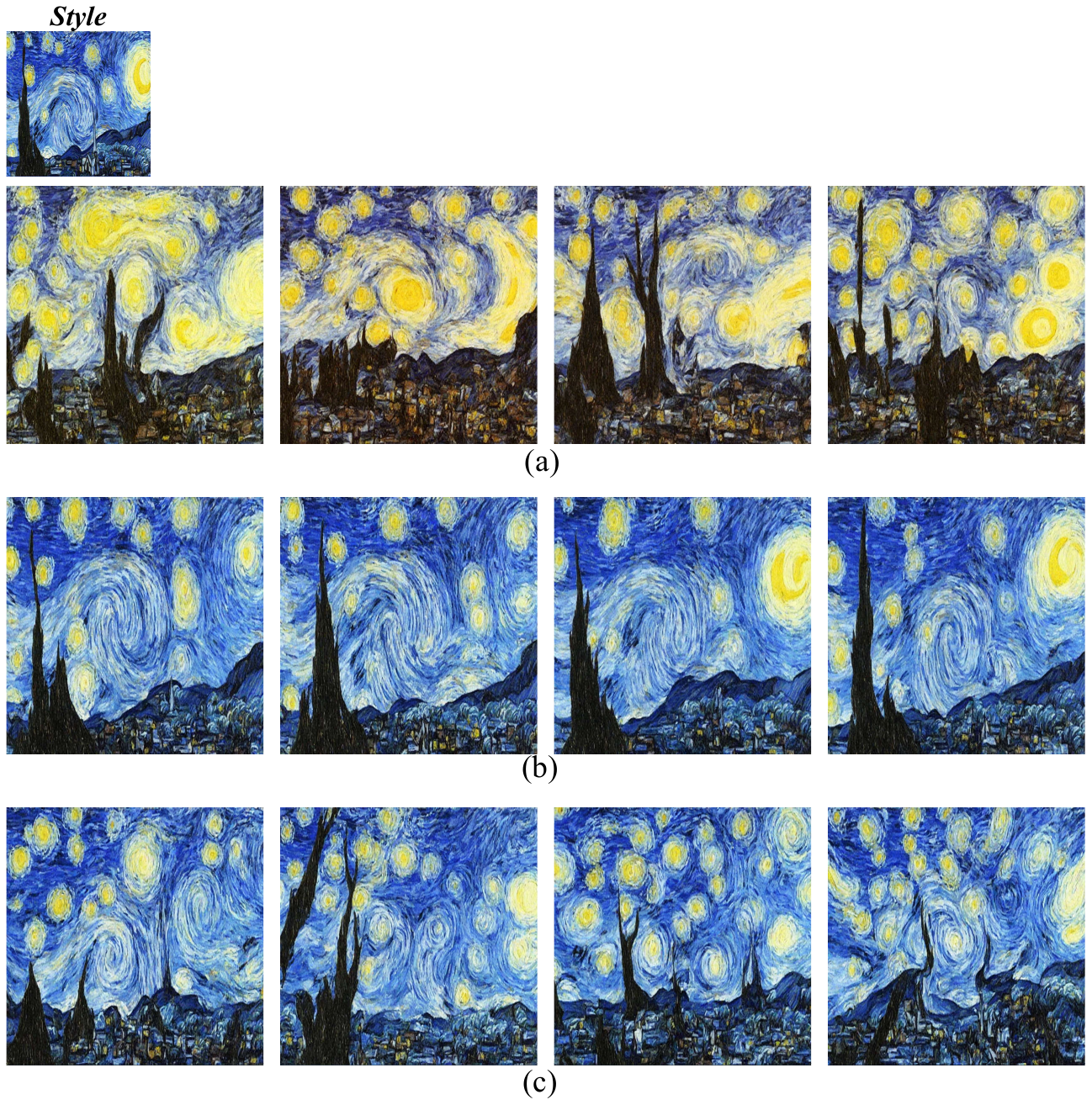}
		\caption{Ablation study on AdaIN and attention layers in SPG sampling. (a) SPG sampling without AdaIN; (b) SPG sampling with KV injection for all self-attention layers; (c) The standard SPG sampling of our method. }
		\label{ablation}
	\end{figure}
	\begin{figure}[t]
		\centering
		\includegraphics[width=\linewidth]{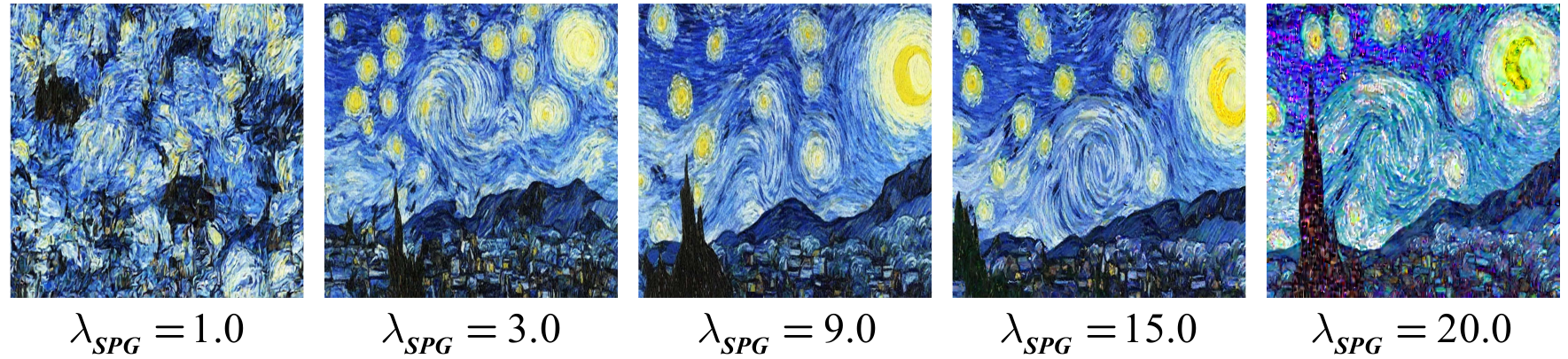}
		\caption{Effect of varying the strength of SPG guidance. }
		\label{parameter}
	\end{figure}
	\subsection{Ablation study}
	
	We first evaluate the impact of AdaIn on SPG sampling. As shown in Fig.~\ref{ablation}(a), disabling AdaIN makes the resulting images completely fail to capture the color distribution of the style reference. In contrast, under the standard SPG setting presented in Fig.~\ref{ablation}(c), the outputs successfully reproduce the correct color tone and detailed style characteristics of the reference, which means incorporating AdaIN during sampling enables a more complete alignment with the given style distribution. Furthermore, we also remove the use of AdaIN in the integration of SPG with CFG. As illustrated by the orange point in Fig.~\ref{point}, the generated images show inferior stylization compared to the standard configuration.

	Then, we investigate the impact of self-attention layers involved in feature injection.
	Intuitively, since KV features encode a substantial amount of style and texture information, one might expect that injecting more would yield better stylization. However, KV also contains global structure information of the reference image. As shown in Fig.~\ref{ablation}(b), the generation process tends to fully reconstruct the original reference image when all self-attention layers (layers 1 to 70) are involved for injection. This ablation result is consistent with the previous empirical design.
	
	We also studied the effect of $\lambda_{\text{SPG}}$. As shown in Fig.~\ref{parameter}, when $\lambda_{\text{SPG}}$ ranges from 3 to 15, SPG still vividly preserves the style features of the reference image, demonstrating strong robustness in choosing $\lambda_{\text{SPG}}$. In our experiments, we set $\lambda_{\text{SPG}}$ to 3 by default.
	
	\begin{figure}[t]
		\centering
		\includegraphics[width=\linewidth]{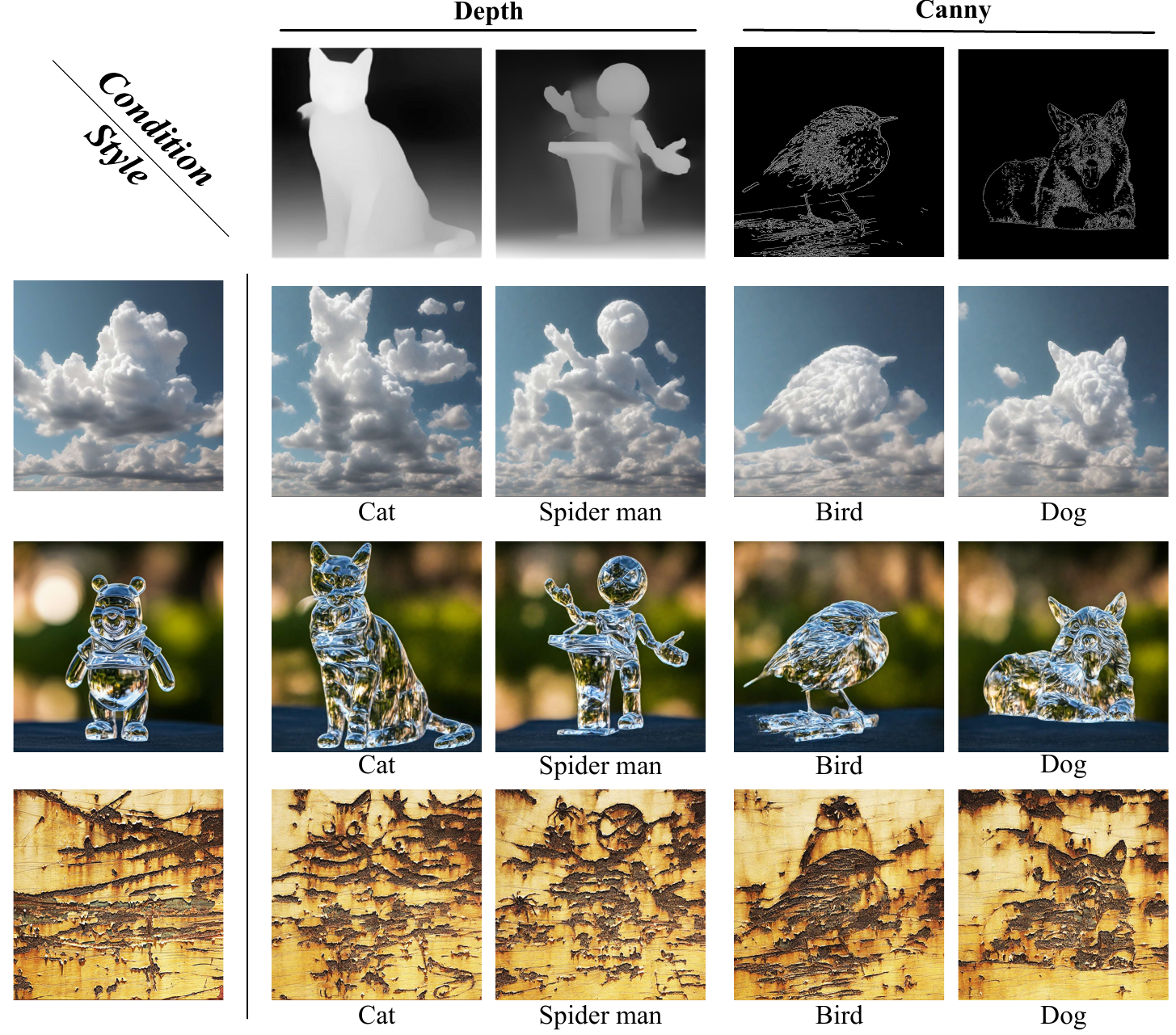}
		\caption{Applying SPG-CFG integration with ControlNet.}
		\label{fig:controlnet}
	\end{figure}
	
	\begin{figure}[t]
		\centering
		\includegraphics[width=\linewidth]{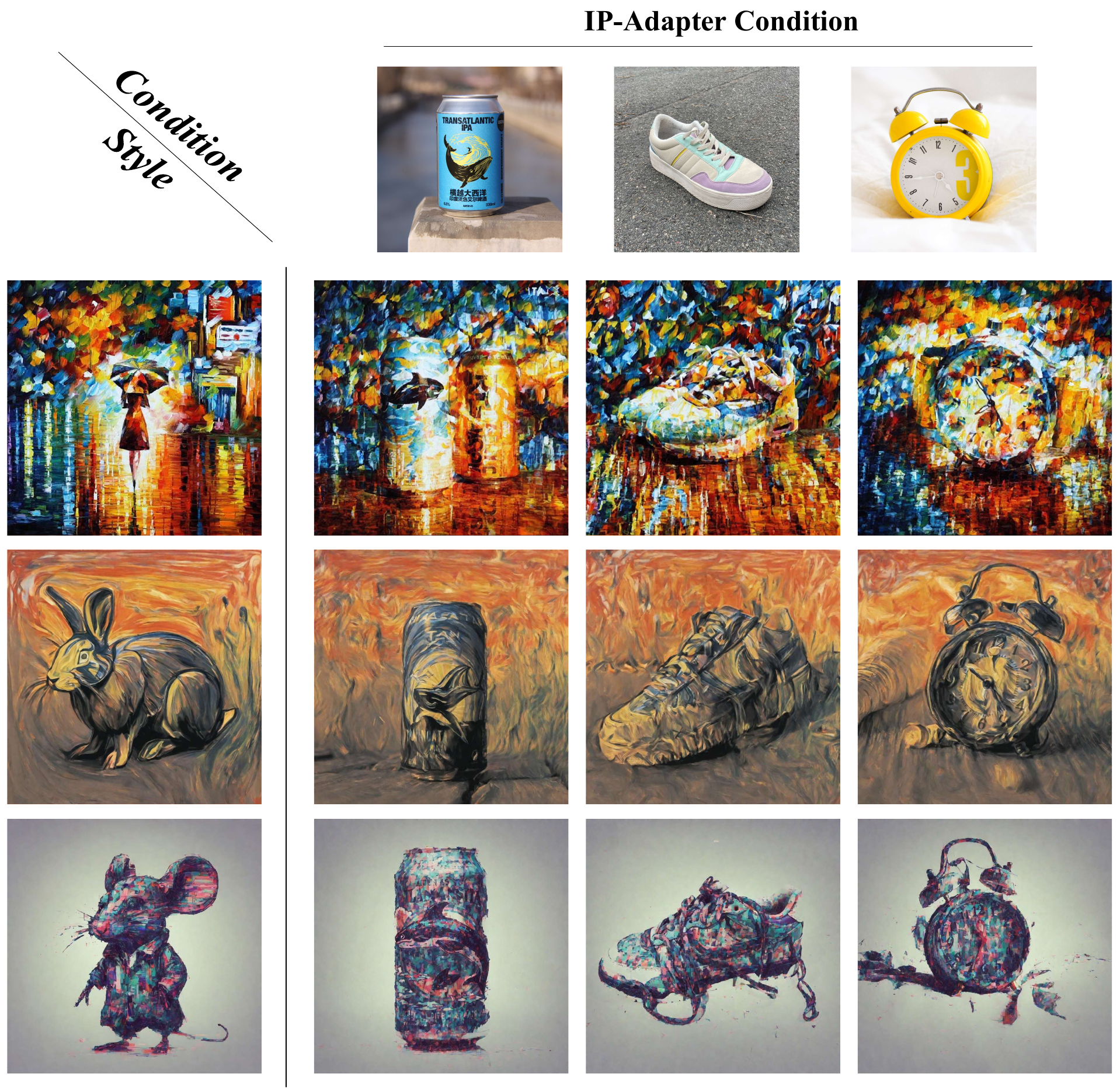}
		\caption{Applying SPG-CFG integration with IP-Adapter.}
		\label{fig:ipadapter}
	\end{figure}
	
	\subsection{Additional results and applications}
	Since SPG is a training-free method, requiring no additional learning, it can be easily integrated with other diffusion-based approaches. As shown in Fig.~\ref{fig:controlnet}, we combine our method with ControlNet~\cite{zhang2023controlnet} by introducing additional controls, \eg Canny edges or depth image, into the existing SPG-CFG framework. Our method is capable of generating images that simultaneously adhere to the reference style, textual prompt, and additional conditions from ControlNet, while exhibiting strong robustness. 
	Fig.~\ref{fig:ipadapter} further shows results integrating our method with IP-Adapter~\cite{ye2023ipadapter}, where image prompts are input as content guidance rather than texts. The generated images successfully mimic the reference style while preserving the identity of the input object.
	
\begin{figure}[tbp]
		\centering
		\includegraphics[width=\linewidth]{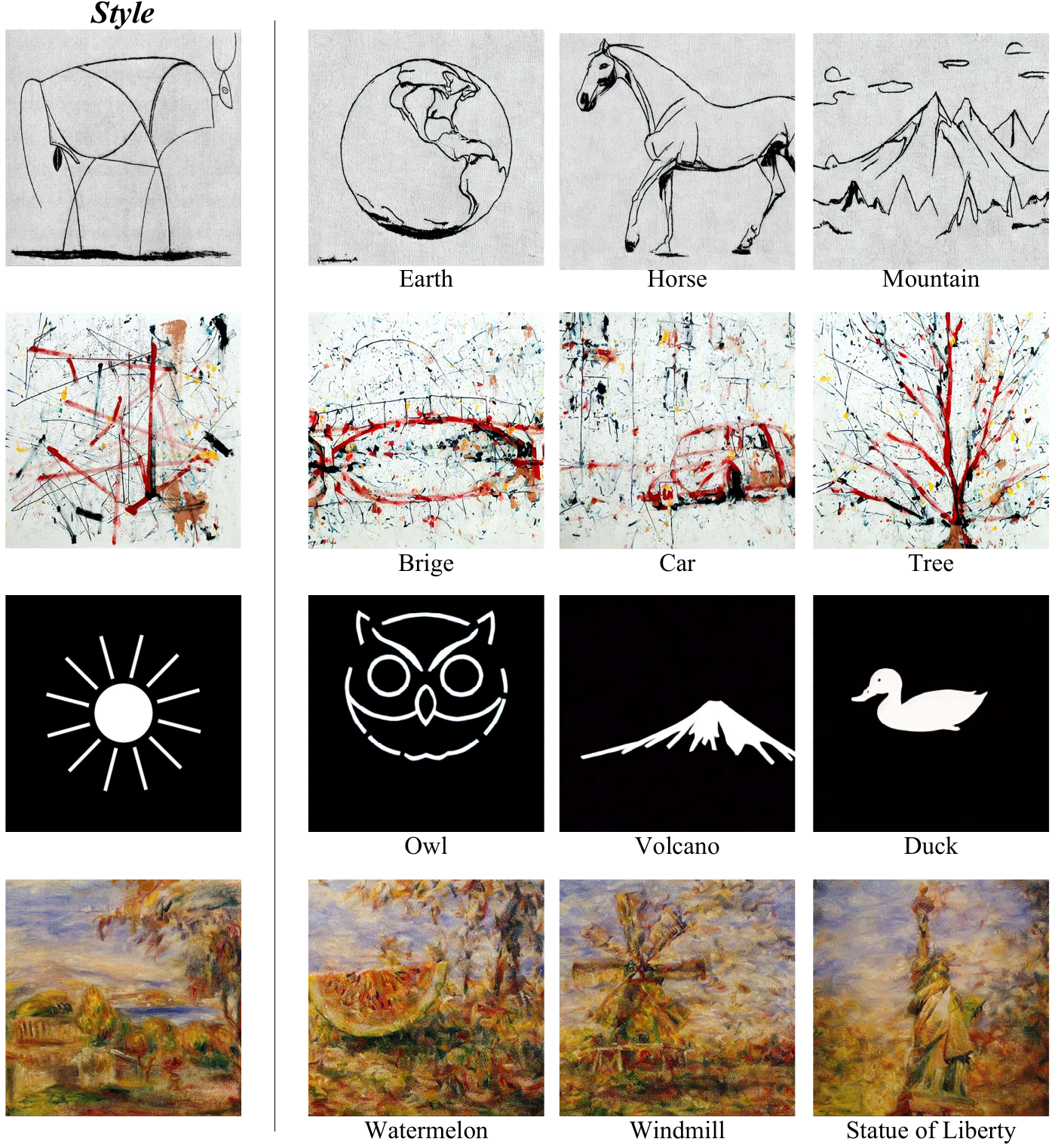}
		\caption{\label{sd1.5} Results obtained on Stable Diffusion v1.5.}
	\end{figure}

	SPG can also be applied to other diffusion models that employ CFG or similar guidance strategies to steer the sampling process toward a desired distribution. As shown in Fig.~\ref{sd1.5}, we apply the new SPG-CFG framework to Stable Diffusion v1.5~\cite{rombach2022highlatentdiffusion} model and still acquire reasonably high-quality stylistic T2I generation results. As for computational cost, generating a stylized image using SD v.15 takes just 7 seconds on an RTX 3090, while still achieving very high-quality results.
	
	\section{Conclusions}
	We propose SPG, a novel sampling method for stylized image generation based on pretrained T2I diffusion models. Unlike existing training-free approaches that typically inject KV features of the reference image into the CFG sampling process, SPG is totally independent of the original CFG, enabling our method to generate stylized images without compromising the strengths of CFG, and thus reaching the best balance between style preservation and semantic alignment. Our method also supports a wide range of diffusion model plugins, demonstrating strong robustness and high quality with various conditional guidance.

	A limitation of our method is its inapplicability to recently introduced models, such as Flux~\cite{flux2024}, which are distilled models prioritizing sampling efficiency through direct noise prediction, rather than CFG or similar guidance mechanisms. 
	We leave the exploration in FLUX as future work.

	\section*{Acknowledgments}
	This work was supported in parts by National Key R\&D Program of China (2024YFB3908500, 2024YFB3908502, 2024YFB390 8505), ICFCRT (W2441020), Guangdong Basic and Applied Basic Research Foundation (2023B1515120026), DEGP Innovation Team (2022KCXTD025), and Scientific Development Funds from Shenzhen University.
	
	\bibliographystyle{eg-alpha-doi} 
	\bibliography{egbibsample}       
	
	
	\newpage

\end{document}